# Solid-State Thermal Energy Storage Using Reversible Martensitic Transformations

***AUTHORS:*** *Darin J. Sharar[1*], Brian F. Donovan[2], Ronald J. Warzoha[2], Adam A. Wilson[3], and Asher C. Leff[4]*

**AFFILIATIONS:**

[1]U.S. Army Research Laboratory, Adelphi, MD 20783, United States

[2]U.S. Naval Academy, Annapolis, MD 21402, United States

[3]National Academy of Sciences, National Research Council, Washington, DC 20001

[4]General Technical Services under contract with U.S. Army Research Laboratory, Adelphi, MD 20783, United States

**CORRESPONDING AUTHOR:** *Correspondence and requests for materials should be addressed to D. Sharar (darin.j.sharar.civ@mail.mil)

**ABSTRACT**

The identification and use of reversible Martensitic transformations, typically described as shape memory transformations, as a new class of solid-solid phase change material is experimentally demonstrated here for the first time. To prove this claim, time-domain thermoreflectance, frequency-domain thermoreflectance, and differential scanning calorimetry studies were conducted on commercial NiTi alloys to quantify thermal conductivity and latent heat. Additional Joule-heating experiments demonstrate successful temperature leveling during transient heating and cooling in a simulated environment. Compared to standard solid-solid materials and solid-liquid paraffin, these experimental results show that shape memory alloys provide up to a two order of magnitude higher Figure of Merit. Beyond these novel experimental results, a comprehensive review of >75 binary NiTi and NiTi-based ternary and quaternary alloys in the literature shows that shape memory alloys can be tuned in a wide range of transformation temperatures (from -50 to 330°C), latent heats (from 9.1 to 35.1 J/g), and thermal conductivities (from 15.6 to 28 W/m·K). This can be accomplished by changing the Ni and Ti balance, introducing trace elements, and/or by thermomechanical processing. Combining excellent corrosion resistance, formability, high strength and ductility, high thermal performance, and tunability, SMAs represent an exceptional phase change material that circumvents many of the scientific and engineering challenges hindering progress in this field.

**MAIN TEXT**

Thermal energy storage (TES) using phase change materials (PCMs) offers tremendous benefits in a diverse array of technology spaces, ranging from large scale power generation to more localized electronic thermal management.[1] A key challenge to TES, and an area of extensive research interest, is synthesis and integration of high performance PCMs. Often-times, solid-liquid phase change materials (SL-PCMs) require engineering measures in the form of encapsulants and metallic fin structures to provide mechanical support, prevent liquid phase PCM leakage, and enhance poor PCM thermal conductivity. In this regard, solid-solid phase change materials (SS-PCMs) have received increased attention because of their intrinsic packaging benefits over more-traditional solid-liquid PCMs.[2-4] However, due to poor thermal conductivity, SS-PCMs still require engineering measures to improve thermal performance. In both cases, standard engineering approaches impose significant design restrictions and reduce volumetric latent heat storage.

We report here for the first time the use of NiTi and NiTi-based shape memory alloy smart materials (SMAs) as ultrahigh performance SS-PCMs. Traditionally, NiTi SMAs have been used in commercial and benchtop applications aimed to exploit their unique strain-induced superelastic/elastocaloric or temperature-induced shape memory properties.[5,6] Both of these effects, and the new PCM response described herein, rely on the solid-solid crystalline phase transformation from the low-temperature monoclinic B19' Martensite (M) structure to the high-temperature cubic B2 Austenite (A) lattice. In some cases, intermediate rhombohedral (R-phase) and orthorhombic B19 phases will form during this process.[7] In addition to the Martensitic phase transformation, NiTi alloys have a number of attractive properties, including excellent corrosion resistance[8], high strength and ductility[9], and good formability via traditional thermomechanical processing, machining, and additive manufacturing.[10]

Two 1 mm thick commercial NiTi alloy plates with different transformation temperatures were purchased to demonstrate the ability of SMAs to function as SS-PCMs. One material was Martensite at room temperature (RTM) and the other was in the high temperature Austenite phase at room temperature (RTA). The RTA sample served as a control with near-identical thermal diffusivity, but lacking a phase transformation when heated above ambient temperature. All material was heat treated at 500 °C for 15 minutes in air and water quenched prior to characterization.

The thermal conductivity and volumetric heat capacity of the material were determined using both time-domain thermoreflectance (TDTR) and frequency-domain thermoreflectance (FDTR). The TDTR system utilized an ultra-fast optical pulse train emanating from a Ti:sapphire oscillator. The oscillator signal was split to a pump beam of 8 MHz amplitude modulated light (frequency doubled to 404 nm) and a probe beam (at 808 nm) that was used to lock in to the pump heating at the sample. The pump optical energy was converted to thermal energy with a PVD E-beam evaporated 80 nm Au transducer layer on top of the SMA. The temperature-dependent change in reflectance of the Au transducer was monitored by the probe beam, which was delayed in arrival time compared to the pump. This time delay resulted in a temporal evolution of the SMA temperature as it underwent the pump heating. The FDTR system utilized a 532 nm pump beam to establish a heating event on the sample surface. The pump was amplitude modulated between 3 kHz and 30 MHz to achieve a variation in the thermal penetration depth within the sample. A separate 808 nm probe beam was used to monitor the change in reflectivity, and thus temperature, as a function of frequency.

For both techniques the recorded data was then fitted against an analytical model to determine thermal conductivity and heat capacity values. A complete description of these techniques can be found in articles by Cahill[11] and Schmidt et al.[12]. The fittings for the data recorded herein can be found in the supplemental material. The combination of both TDTR and FDTR provided increased fidelity of measured values, as well as determination of multiple thermophysical parameters.

Differential Scanning Calorimetry (DSC) was carried out for both RTA and RTM materials using a Perkin-Elmer 8500 DSC. Prior to DSC, the specimens were submerged in liquid nitrogen and then allowed to return to ambient temperature to ensure that they were on the heating side of the phase transformation hysteresis loop. Samples were stabilized for 5 minutes to within $\pm$0.1°C of the setpoint start temperature of 26°C. Then, samples were ramped at a rate of 10°C/min to the maximum temperature of 70°C, allowed to dwell for 5 minutes, then cooled at a rate of 10°C/min back to 26°C.[13] Baseline subtraction for the Aluminum test pans was performed.

25.4 mm x 25.4 mm samples of both RTA and RTM materials were also characterized via constant power Joule-heating at 3W. ATC 50-ohm ceramic chip resistors were attached directly to the specimens and heated for two minutes followed by passive cooling. Throughout this process the temperature of the samples was monitored using a FLIR A40 thermal camera with a precision of $\pm$ 1°C. Finally, the density of the material was verified using mass and fluid displacement.

The DSC performed on the RTA specimen returned a constant heat flow versus temperature curve, confirming that no phase transformation occurs during heating in the temperature regime being investigated. The DSC results for the RTM specimen, shown in Figure 1(a), revealed the Martensite to Austenite transition at a peak transformation temperature of 57°C. The measured endothermic latent heat was 13.46 J/g. Upon cooling, the reverse Austenite to intermediate R-phase transformation occurred at a temperature of 35°C, resulting in a latent heat release of 4.6 J/g.

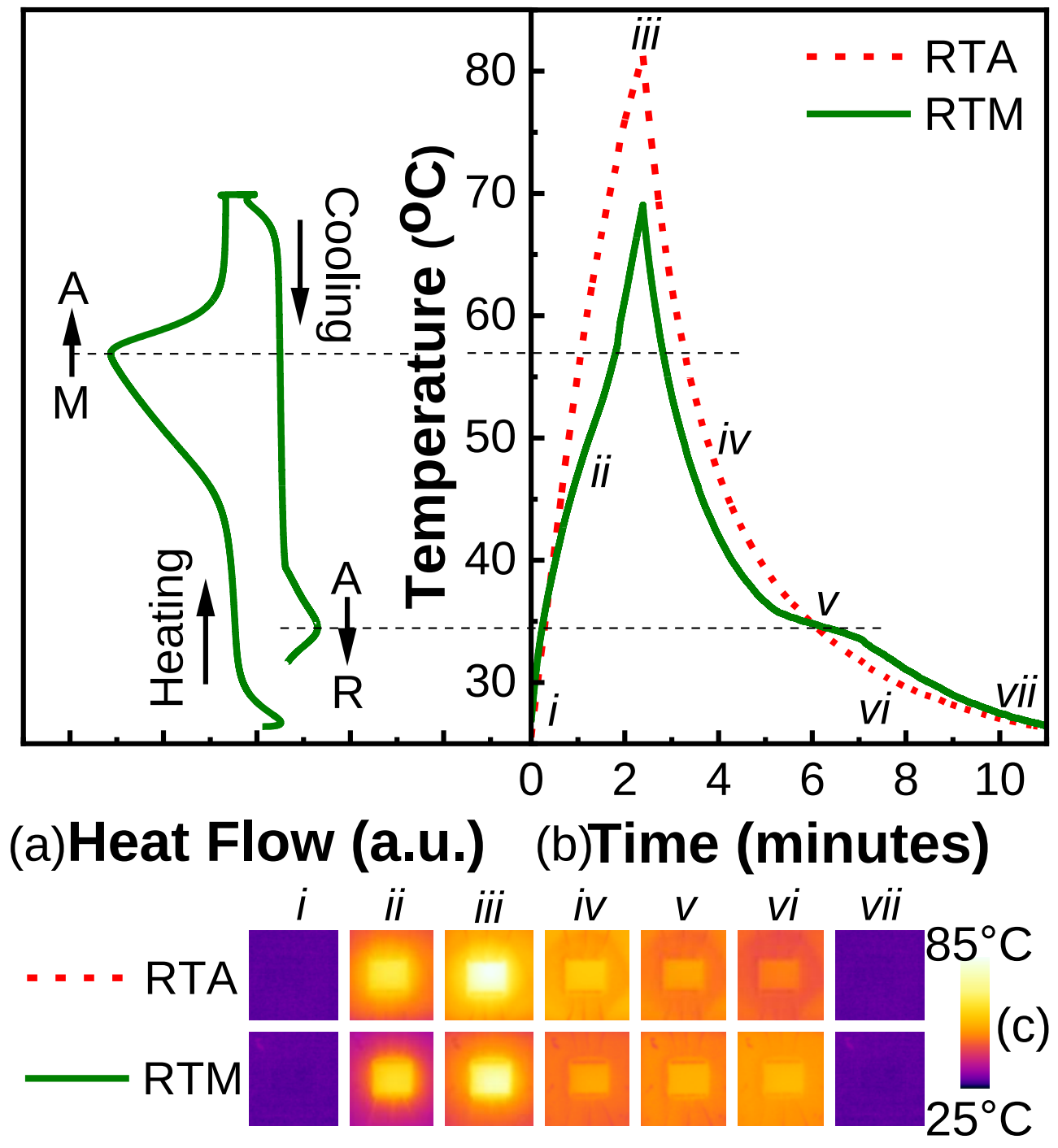

**Figure 1: (a) Differential Scanning Calorimetry result for the NiTi alloy used in the current study. Corresponding thermally-activated phase transformations are indicated. (b) Thermal response of the RTM alloy (solid-green) and a benchmark RTA alloy (dashed-red) during Joule heating. (c) IR camera images captured at instants corresponding to locations i-vii on 1(b).**

The results of the constant power Joule-heating experiment are shown in Figures 1(b) and 1(c). Initially, both samples started at room temperature (25°C) and embarked on rapid sensible heating once the power was supplied. An inflection point was observed in the RTM sample at a temperature of 45°C, corresponding with the start of the M→A solid-solid phase transformation. Another inflection point can be seen at 57°C, corresponding with the peak of the DSC heating curve. The M→A phase transformation exhausted shortly after this peak and the RTM temperature began tracking parallel to the RTA curve as both materials sensibly heated. The RTA sample reached a maximum temperature of 82°C while the RTM sample did not exceed 68°C. This represents 14°C cooling, or a 25% reduction in the maximum temperature rise.

Upon passive cooling the RTA material responded in a characteristic exponential decay to room temperature. The RTM material sensibly cooled from 68°C to 40°C, then underwent the reverse A→R transformation between 40°C and 35°C. This resulted in a clear non-linear temperature response and, briefly, a hotter RTM material. The RTM and RTA samples both recovered to room temperature at the end of the experiment.

It's clear from the above side-by-side analysis, Figure 1(a) and (b), that the non-linear temperature response during Joule-heating corresponds to endo- and exotherms in the DSC data. This proves that the temperature response during heating and cooling is a result of the observed phase transformations. The Joule-heating experiment also serves as a direct demonstration of temperature leveling during a transient heating and cooling cycle.

The above experimental analysis is useful for illustrative purposes; however, the absolute temperatures, time scales, and percent improvements described here are unique to the arbitrary heating and cooling conditions, ambient conditions, sample size, and parasitic losses. Extrapolating these results to more massive or energy dense systems would be unreliable. Instead, the FOM described by T.J. Lu is a more appropriate quantifiable measure of PCM performance.[14] Lu defines a PCM FOM as:

$$FOM = \rho \times L \times k_{HT} \tag{1}$$

Where ρ is density, L is the latent heat of transformation, and $k_{HT}$ is the thermal conductivity of the high temperature phase. Thus, a high FOM represents a high volumetric latent heat capacity and the ability to readily charge and discharge thermal energy. The FDTR system returned a thermal conductivity of 17.3 $Wm^{-1}K^{-1}$ for the RTM material when measured at 80 °C. For the same sample, the TDTR tool produced a thermal conductivity of 17.9 $Wm^{-1}K^{-1}$ when taken at 70.5 °C. The density of the RTM material was measured to be 6,239 $\pm$ 3% $kg/m^3$, which is in good agreement with the literature.[15] The calculated FOM for the NiTi RTM sample in this study was $1{,}478\times10^{3}$ $J^2K^{-1}s^{-1}m^{-4}$ using the mean of the two values for the thermal conductivity (17.6 $Wm^{-1}K^{-1}$), combined with the measured density and latent heat. This FOM value is reported in Figure 2.

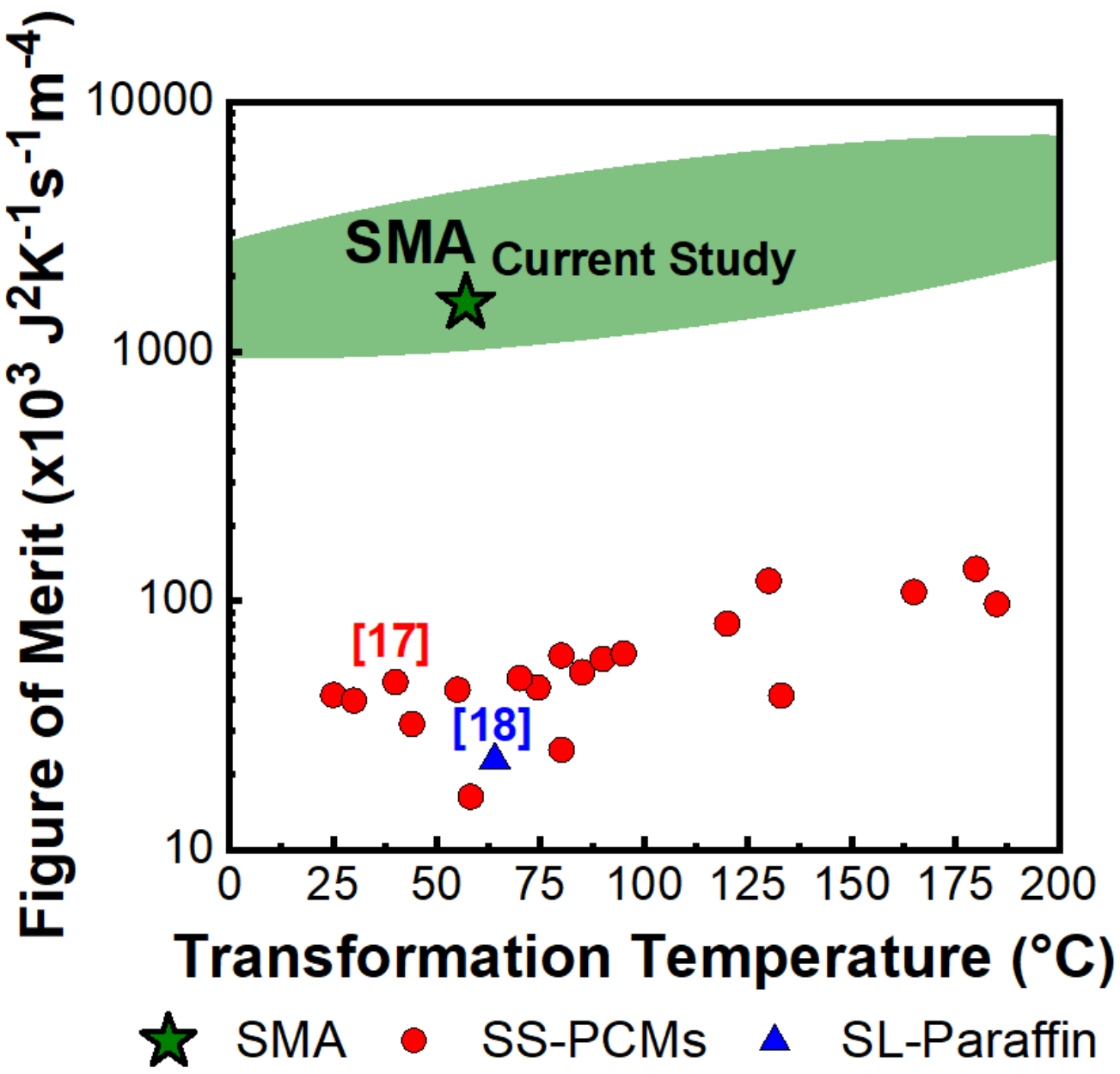

**Figure 2: TES Figure of Merit versus transformation temperature for the RTM NiTi characterized herein (green star), along with values for other SS-PCMs and paraffin, which is a SL-PCM. A green band is included showing the potential range of FOM and transformation temperature for NiTi based alloys based on thermal conductivity, density, and latent heat values from the literature.**

Transformation temperatures ranging from -50 to 330°C, latent heats from 9.1 to 35.1 J/g, and Austenite thermal conductivities from 15.6 to 28 W/m·K have been reported for NiTi-based materials in the literature. These properties can be tuned by altering the Ni and Ti balance, introducing trace elements, and/or by thermomechanical processing.[16] The green band in Figure 2 represents the range of FOM expected based on these reported latent heat values, transition temperatures, and the upper and lower thermal conductivity bounds. These results are plotted along with representative SS-PCM[17] (FOM=$43\times10^{3}$ $J^2K^{-1}s^{-1}m^{-4}$) and paraffin[18] (FOM=$23\times10^{3}$ $J^2K^{-1}s^{-1}m^{-4}$) values on Figure 2 and are summarized in Table S1 in the supplemental material. As shown, NiTi-based alloys, including the commercial material we tested herein, offer up to two orders of magnitude higher FOM relative to standard SS-PCMs and paraffin.

This result is illustrated graphically in a multi-parameter radar plot shown in Figure 3, where the size of the shaded area for each technology represents the FOM. As shown, NiTi has comparable volumetric latent heat ($\rho \times L$) with significantly higher thermal conductivity (k).

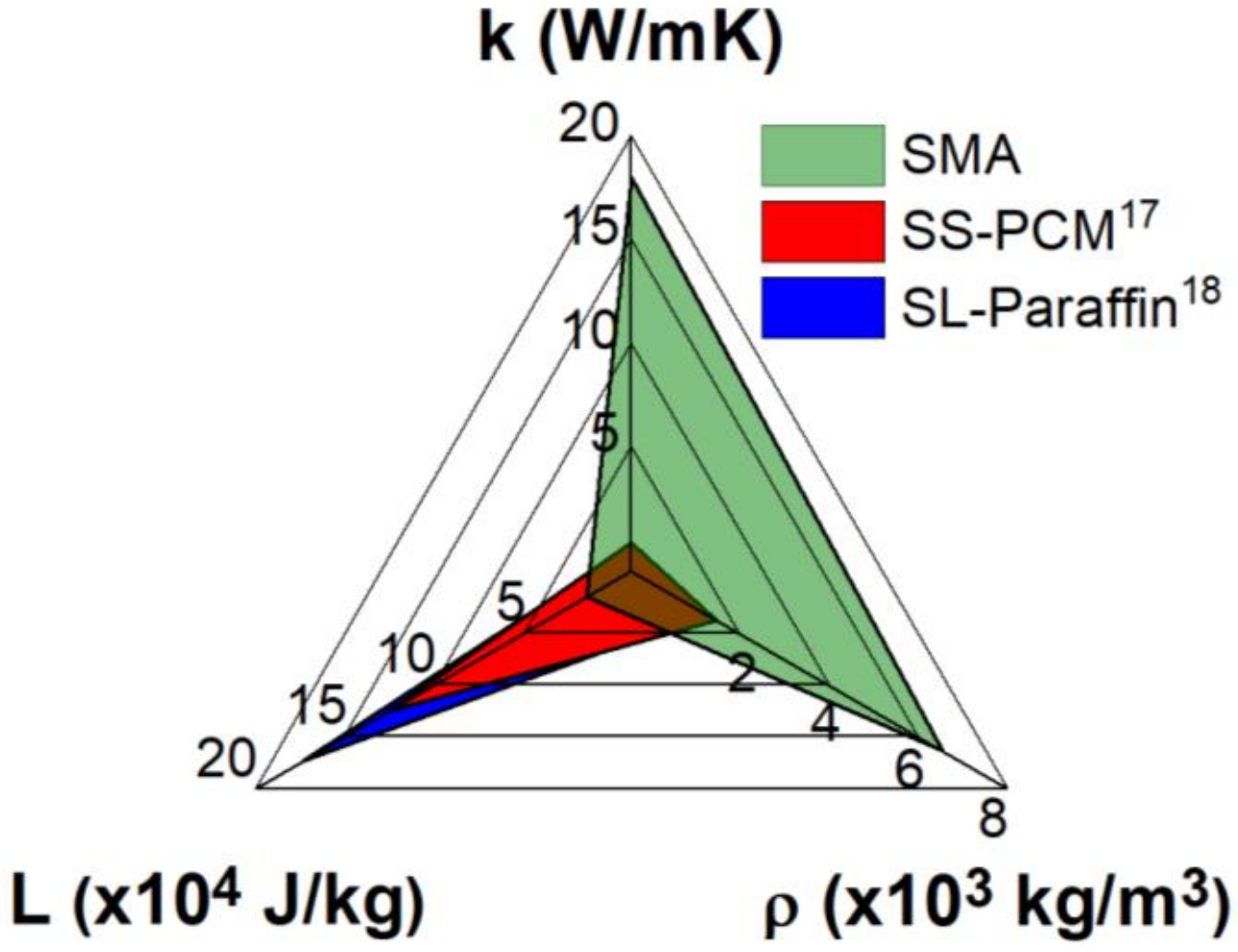

**Figure 3: Radar plot with axes for latent heat (L), density (ρ), and thermal conductivity (k) with a green area showing the properties for the RTM NiTi measured herein, a red area corresponding to the SS-PCM values[17] and a blue area corresponding to the values for paraffin[18].**

In conclusion, it has been identified and demonstrated here that reversible Martensitic transformations, typically described as shape memory transformations, can be utilized for thermal energy storage. This newly-identified class of solid-solid PCM shows a two order of magnitude FOM improvement over traditional PCMs. In addition to favorable performance, the SMA transformation temperature, latent heat, thermal conductivity, and structural properties can be easily tuned by altering the Ni and Ti balance, introducing trace elements, and adjusting thermomechanical processing. These favorable properties and attributes, along with their corrosion resistance and manufacturability, will allow the community to pivot away from the standard fin-, encapsulant-, or additive-enhanced archetype and begin fabricating multifunctional structural, heat transfer, and active energy absorber/discharger components from a single SMA material. The breakthrough discovery here of SMAs as a new class of high performance, tunable PCM presents an opportunity for a paradigm shift in the community's development of high-performance TES systems.

## ASSOCIATED CONTENT

NiTi sample preparation, Joule-heating thermal characterization, time-domain thermoreflectance (TDTR) and frequency-domain thermoreflectance (FDTR), determination of material density, elemental analysis using energy-dispersive X-ray spectroscopy (EDS), data reduction for calculating FOM, and Table S1 which lists materials, compositions, transition temperatures, latent heat, and FOM

## AUTHOR INFORMATION

*E-mail: darin.j.sharar.civ@mail.mil

## ACKNOWLEDGMENT

RJW gratefully acknowledges support from the Office of Naval Research and Dr. Mark Spector under contract number N0001418WX00343. RJW and BFD also gratefully acknowledge support from the Office of Naval Research and Mr. Peter Morrison under contract number N0001418WX00039-1. A portion of this research was performed while AAW held an NRC Research Associateship award at the U.S. Army Research Laboratory.

## NOTES

The authors declare no competing financial interest.

## REFERENCES

(1) Nazir, H.; Batool, M.; Bolivar Osorio, F.; Isaza-Ruiz, M.; Xu, X.; Vignarooban, K.; Phelan, P.; Inamuddin; Kannan, A. Recent developments in phase change materials for energy storage applications: A review. *Int. J. of Heat and Mass Trans.* **2019**, *129, 491-523*.

(2) Jankowski, N.; McCluskey, F. A review of phase change materials for vehicle component thermal buffering. *App. Energy.* **2014**, *113, 1525-1561.*

(3) Fallahi, A.; Guldentops, G.; Tao, M.; Granado-Focil, S.; Van Dessel, S. Review on solid-solid phase change materials for thermal energy storage: Molecular structure and thermal properties. *App. Therm. Eng.* **2017**, *127, 1427-1441.*

(4) Farid, M.; Khudhair, A.; Razak, S.; Al-Hallaj, S. A review on phase change energy storage: Materials and applications. *Energy Conv. and Manag.* **2004**, *45, 1597-1615.*

(5) Cui, J.; Wu, Y.; Muehlbauer, J.; Hwang, Y.; Radermacher, R.; Fackler, S.; Wuttig, M.; Takeuchi, I. Demonstration of high efficiency elastocaloric cooling with large ΔT using NiTi wires. *App. Phys. Lett.* **2012**, *101, 1078-1113.*

(6) Jani, J.; Leary, M.; Subic, A.; Gibson, M. A review of shape memory alloy research, applications, and opportunities. *Mater. and Design.* **2014**, *56, 1078-1113.*

(7) Frenzel, J.; Wieczorek, A.; Opahle, I.; MaaB, B.; Drautz, R.; Eggeler, G. On the effect of alloy composition on martensite start temperatures and latent heats in Ni-Ti based shape memory alloys. *Acta Mater.* **2015**, *90, 213-231.*

(8) Rondelli, G. Corrosion resistance tests on NiTi shape memory alloy. Biomaterials. **1996**, 17, 2003-2008.

(9) Casalena, L.; Bucsek, A.; Pagan, D.; Hommer, G.; Bigelow, G.; Obstalecki, M.; Noebe, R.; Mills, M.; Stebner, A. Structure-property relationships of a high strength superelastic NiTi-1Hf alloy, *Adv. Eng. Mater.* **2018**, 20, 1800046.

(10) Hou, H.; Simsek, E.; Stasak, D.; Hasan, N.; Qian, S.; Ott, R.; Cui, J.; Takeuchi, I. Elastocaloric cooling of additive manufactured shape memory alloys with large latent heat. *J. of Phys. D App. Phys.*, **2017**, 50, 10.1088

(11) Cahill, D.G.; Analysis of heat flow in layered structures for time-domain thermoreflectance. *Review of Scientific Instruments*. **2004**, 75, 5119-5122.

(12) Schmidt, A.J.; Cheaito, R.; Chiesa, M. A frequency-domain thermoreflectance method for the characterization of thermal properties. *Review of Scientific Instruments*. **2009**, 80, 094901.

(13) ASTM C1784-14: Standard test method for using a heat flow meter apparatus for measuring thermal storage properties. *ASTM International, West Conshohocken, PA*, **2014**.

(14) Lu, T.J. Thermal management of high power electronics with phase change cooling. *Int. J. of Heat and Mass Trans.* **2000**, *43, 2245-2256.*

(15) Boyer, R.; Collings, E.W.; Welsch, G. *Materials Properties Handbook: Titanium Alloys*. ASM International, Materials Park, USA, **1994**.

(16) Shen, Y.; Zhou, H.; Zheng, Y.; Peng, B.; Haapasalo, M.; Current challenges and concepts of the thermomechanical treatment of nickel-titanium instruments. JOE, 2013, 39, 163-172.

(17) P.P. Limited. PlusICE Phase Change Materials. [online] available: http://www.pcmproducts.net/files/PlusICE%20Range-2013.pdf.

(18) Shao, L.; Raghavan, A.; Kim, G.; Emurian, L. Figure of merit for phase change materials used in thermal management. *Int. J. of Heat and Mass Trans.* **2016**, *101, 764-771*.

# Supporting Information for:
# Solid-State Thermal Energy Storage Using Reversible Martensitic Transformations

***AUTHORS:*** *Darin J. Sharar[1*], Brian F. Donovan[2], Ronald J. Warzoha[2], Adam A. Wilson[3], and Asher C. Leff[4]*

**AFFILIATIONS:**

[1]U.S. Army Research Laboratory, Adelphi, MD 20783, United States

[2]U.S. Naval Academy, Annapolis, MD 21402, United States

[3]National Academy of Sciences, National Research Council, Washington, DC 20001

[4]General Technical Services under contract with U.S. Army Research Laboratory, Adelphi, MD 20783, United States

**CORRESPONDING AUTHOR:** *Correspondence and requests for materials should be addressed to D. Sharar (darin.j.sharar.civ@mail.mil)

**NiTi sample preparation**

RTA and RTM NiTi samples were purchased from a commercial vendor in sheet form and cut into 25.4 x 25.4 mm samples. The samples were 1mm thick. After cutting, the materials were heat treated at 500°C for 15 minutes in air then water quenched. Some heat treated samples were cut into roughly 3.5 x 3.5 mm chips for DSC testing, TDTR testing, FDTR testing, density measurements, and EDS. Others were left in larger form factors for Joule-heating thermal characterization.

**Joule-heating thermal characterization**

The 25.4 x 25.4 mm samples were mechanically polished, then cleaned with a solvent rinse. Next, ATC 50-ohm ceramic chip resistors were attached to the top side of the samples with EPO-TEK P1011 epoxy. A test fixture for holding the samples was made out of polycarbonate material using a Stratasys FDM Titan. Electrical connections to the chip resistors were made using an Aluminum wire bonder to adjacent copper leads. After assembly, the top of the samples (containing the chip resistors) were sprayed with Boron Nitride to provide uniform emissivity. During testing, electrical power was supplied using an Agilent 5951 DC power supply. A FLIR A40 thermal camera was used to monitor the sample temperature (± 1°C accuracy). These results are represented in Figure 1b-c.

**Time-Domain Thermoreflectance (TDTR) and Frequency-Domain Thermoreflectance (FDTR)**

For this work, both time-domain thermoreflectance (TDTR) and frequency-domain thermoreflectance (FDTR) were used to confirm the thermal conductivity and volumetric heat capacity of the NiTi RTM sample. The TDTR system utilized an ultra-fast optical pulse train emanating from a Ti:sapphire oscillator. The oscillator signal was split to a pump beam of 8 MHz amplitude modulated light (frequency doubled to 404 nm) and a probe beam (at 808 nm) that was used to lock in to the pump heating at the sample. The pump optical energy was converted to thermal energy with a PVD E-beam evaporated 80 nm Au transducer layer on top of the SMA. The reflectivity of the sample was measured via the probe beam, which was time delayed such that it directly monitored the surface reflectivity as heat penetrated inward into the sample. The surface reflectivity is directly related to the complex surface temperature profile of the sample via a multi-layer analytical model that is used to fit the data to the NiTi's thermal conductivity. As shown in Figure S1, the ratio, R, of the real part ($V_{in}$) to the imaginary part ($V_{out}$) as measured at the lock-in amplifier ($R = -V_{in}/V_{out}$) is fit as a function of probe delay to determine thermal conductivity, as described elsewhere. [1]

FDTR is an optical pump-probe technique that monitors the surface temperature of the sample as a function of frequency. Our implementation of FDTR utilizes a 532 nm pump beam to establish a heating event on the sample surface. The pump is modulated between 3 kHz and 30 MHz using an electro-optic modulator (Conoptics Model 350-80) to achieve a variation in the thermal penetration depth within the sample. A separate 808 nm probe beam was used to monitor the change in reflectivity (and thus temperature) as a function of frequency. Here, changes in reflectivity are measured as a phase offset ($\Phi$) from the frequency reference of the pump. The same multilayer analytical model may be used with minor adjustments to fit the phase vs frequency signal for thermal conductivity of the sample.[2] A general explanation of the working principle behind each system can be found elsewhere.[1,2]

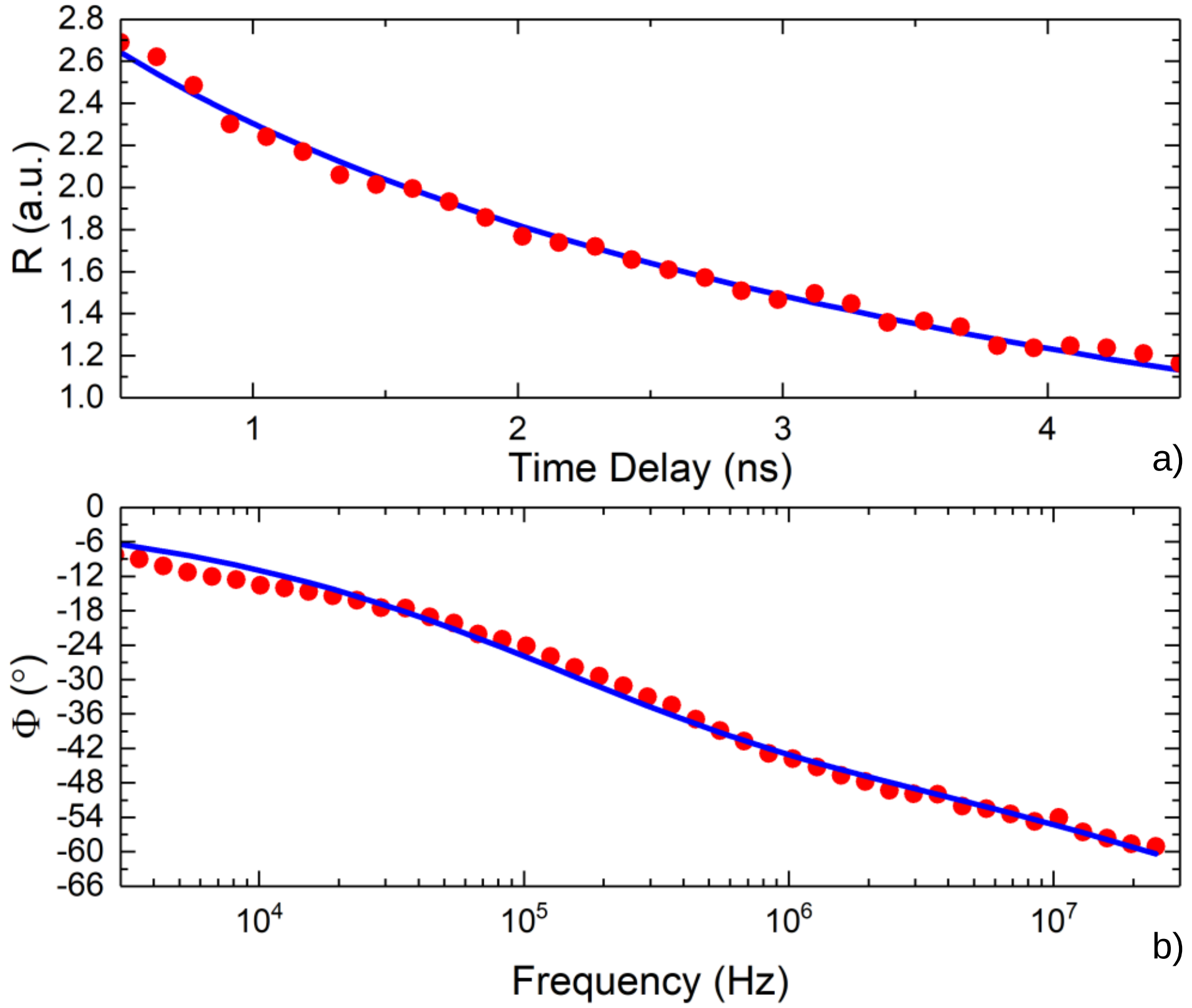

**Figure S4: (a) TDTR ratio ($R=-V_{in}/V_{out}$) vs time delay data (red circles) and model fit (solid blue line) for the RTM sample at 70.5 °C and (b) FDTR phase vs frequency data (red circles) and model fit (solid blue line) for the RTM sample at 80°C. This data was processed via a multi-layer analytical model to calculate a thermal conductivity value between 17.9 and 17.3 W/mK for the TDTR and FDTR systems, respectively.**

For the TDTR system, pump and probe diameters were 60 µm and 12 µm, respectively, while for the FDTR system the pump and probe diameters were 20 µm and 11 µm, respectively. By adjusting the intensity of the pump beam in each system, we measured the thermal conductivity of the NiTi sample at a temperature of 70.5°C via TDTR and 80°C via FDTR. As shown in Figure 1a-b, the material was fully Austenitic at these temperatures. The thermal conductivity using TDTR was calculated to be 17.9 W/m·K and with FDTR was calculated to be 17.3 W/m·K. The average value of 17.6 W/m·K is reported in the main article. The volumetric heat capacity of the sample ($C = \rho \cdot C_p$) was found to be $C = 1.98$ MJ/m$^3$·K using the data obtained with FDTR.

**Determination of material density**

The density of the NiTi material was calculated by dividing the measured mass by fluid displacement. The reported density is 6,239 +3% kg/$m^3$ (6,061 kg/$m^3$ < $\rho$ < 6,428 kg/$m^3$). This reported value is in line with the commonly reported literature value of 6,450 kg/$m^3$.[3]

**Elemental analysis using Energy-Dispersive X-ray Spectroscopy (EDS)**

Elemental analysis was performed using an EDAX Octane Elite EDS with a 70$mm^2$ detector in a Zeiss Auriga SEM environment with an accelerating voltage of 10kV. Samples were prepared by exfoliation, polishing, and solvent rinsing. Energy dispersion spectra were acquired such that the major Ni and Ti peaks each reached >10,000 counts, resulting in 47.73 at.% Ni and 52.27 at.% Ti.

**Data Reduction - Calculating FOM**

Table S1 is a list of material class, composition, low-to-high transition temperature, endothermic latent heat, and associated reference for the materials shown in the Abstract figure. The first entry, $Ni_{47.73}Ti$*, is the material tested in the current study. As described above, the thermal conductivity and density were measured to be 17.6 W/m·K and 6,239 kg/$m^3$, respectively. Combined with a measured latent heat of 13.46 J/g, this yields a FOM value of 1478 x$10^3$ $J^2K^{-1}s^{-1}m^{-4}$.

The thermal conductivity and density values were not originally reported for the NiTi-based alloys listed in Table S1.[4,5] The density of NiTi is a well-known material property that varies negligibly with alloy composition in the range where the B2 Austenite phase is stable. Proper casting/deposition and thermomechanical processing will always result in a non-porous, fully dense part. Based on this understanding, and the lack of reported density values in the literature, a widely-cited[3] value of 6,450 kg/m$^3$ was used to calculate the FOM for the extent of NiTi-based alloys in Table S1 (excluding the current study results).

The thermal conductivity of these NiTi-based metals is expected to be dependent on a number of factors including the crystal structure, alloy composition, phase, grain size, dislocation density, and measurement temperature. Despite the expected variations, there are general trends and experimental ranges reported in the literature. In general, the low temperature B19' Martensite phase has a lower thermal conductivity than the high temperature B2 Austenite phase and both vary based on the exact composition and thermomechanical processing. More specifically, according to Chirtoc et al. the thermal conductivity for B19' Martensite in NiTi ranges from 8.6 to 14 W/m·K while thermal conductivity for B2 Austenite ranges from 15.6 to 28 W/m·K.[6] To give an accurate representation of the available data, the FOM values in Table S1 have been presented as a range calculated with the high and low reported Austenite values of 15.6-28 W/m·K. This is appropriate because the FOM is generally calculated based on the thermal conductivity of the high temperature phase, in this case the Austenite phase.[7] The green band on the abstract image reflects these calculated high and low literature values for the range of NiTi materials.

The material properties required to calculate FOM are readily available for standard SS-PCMs and paraffin. The thermal conductivity and density, respectively, are 0.167 W/m·K and 790 kg/m$^3$ for paraffin, 0.23 W/m·K and 960-1390 kg/m$^3$ for SS-PCMs falling between Neopentyl glycol (NPG) and Pentaerythritol (PE), and 0.36 W/m·K and 1046-1330 kg/m$^3$ for the PlusICE materials as per reported literature values and the manufacturer website.[7-9] These values were used to calculate the FOM values shown in Table S1 and the Figure 2 in the main text.

Table S1: Materials, transition temperatures, thermally-induced latent heat, FOM, and source

| **Material Class** | **Composition** | **Transition temperature (°C)** | **Latent heat $L_{M-A}$ (J/g)** | **FOM ($x10^3$ $J^2/(Ksm^4)$)** | **Reference** |
|---|---|---|---|---|---|
| **Shape Memory Alloy** | $Ni_{47.73}Ti$* | 57 | 13.46 | 1478 | Current Study |
| . | $Ni_{48.62}Ti$ | 90 | 27 | 2716-4876 | 4 |
| . | $Ni_{49.5}Ti$ | 91 | 30.1 | 3028-5436 | . |
| . | $Ni_{49.6}Ti$ | -3.8 | 18.3 | 1841-3304 | . |
| . | $Ni_{49.63}Ti$ | 46.7 | 21.6 | 2173-3900 | . |
| . | $Ni_{49.68}Ti$ | 81.9 | 31.7 | 3189-5725 | . |
| . | $Ni_{49.74}Ti$ | 86.3 | 22.9 | 2304-4135 | . |
| . | $Ni_{50.01}Ti$ | 52.7 | 27 | 2716-4876 | . |
| . | $Ni_{50.13}Ti$ | -21.9 | 14.9 | 1499-2690 | . |
| . | $Ni_{50.17}Ti$ | -39.1 | 9.1 | 915-1643 | . |
| . | $Ni_{50.18}Ti$ | 47.2 | 26.7 | 2686-4822 | . |
| . | $Ni_{50.21}Ti$ | 4.1 | 16.5 | 1660-2979 | . |
| . | $Ni_{50.42}Ti$ | 17 | 23.3 | 2344-4207 | . |
| . | $Ni_{50.49}Ti$ | 17.8 | 23.9 | 2404-4316 | . |
| . | $Ni_{50.58}Ti$ | -10.8 | 18.8 | 1891-3395 | . |
| . | $Ni_{50}Ti_{49.9}Cr_{0.1}$ | 89.25 | 30 | 3018-5418 | 5 |
| . | $Ni_{50}Ti_{49.75}Cr_{0.25}$ | 63.25 | 23.7 | 2384-4280 | . |
| . | $Ni_{50}Ti_{49.5}Cr_{0.5}$ | 42.35 | 23.3 | 2344-4207 | . |
| . | $Ni_{50}Ti_{49.35}C_{50.65}$ | 33.2 | 21.3 | 2143-3846 | . |
| . | $Ni_{50}Ti_{49.2}Cr_{0.8}$ | 13.05 | 20.8 | 2092-3756 | . |
| . | $Ni_{50}Ti_{49}Cr_{1}$ | -12.8 | 15.2 | 1529-2745 | . |
| . | $Ni_{50}Ti_{48.75}Cr_{1.25}$ | -46.2 | 10.1 | 1016-1824 | . |
| . | $Ni_{50}Ti_{49.9}V_{0.1}$ | 101.15 | 30.1 | 3028-5436 | . |
| . | $Ni_{50}Ti_{49.75}V_{0.25}$ | 82.9 | 24.9 | 2505-4496 | . |
| . | $Ni_{50}Ti_{49}V_{1}$ | 64.5 | 23.2 | 2334-4189 | . |
| . | $Ni_{50}Ti_{48}V_{2}$ | 54.7 | 22.2 | 2233-4009 | . |
| . | $Ni_{50}Ti_{47}V_{3}$ | 35.9 | 16.9 | 1700-3052 | . |
| . | $Ni_{50}Ti_{45}V_{5}$ | 18.3 | 12 | 1207-2167 | . |
| . | $Ni_{50}Ti_{44}V_{6}$ | 15.15 | 10.6 | 1066-1914 | . |
| . | $Ni_{43}Ti_{52}Cu_{5}$ | 73.2 | 24.2 | 2435-4370 | . |
| . | $Ni_{43.5}Ti_{51.5}Cu_{5}$ | 77.6 | 24.7 | 2485-4460 | . |
| . | $Ni_{44}Ti_{51}Cu_{5}$ | 77.75 | 24.1 | 2424-4352 | . |
| . | $Ni_{44.25}Ti_{50.75}Cu_{5}$ | 73.95 | 25.9 | 2606-4677 | . |
| . | $Ni_{44.5}Ti_{50.5}Cu_{5}$ | 74.75 | 25.7 | 2585-4641 | . |
| . | $Ni_{44.6}Ti_{50.4}Cu_{5}$ | 77.9 | 29.1 | 2928-5255 | . |
| . | $Ni_{44.75}Ti_{50.25}Cu_{5}$ | 81.55 | 23.6 | 2374-4262 | . |
| . | $Ni_{45}Ti_{50}Cu_{5}$ | 80.65 | 29.3 | 2948-5291 | . |
| . | $Ni_{45.2}Ti_{49.8}Cu_{5}$ | 66.5 | 27 | 2716-4876 | . |
| . | $Ni_{45.3}Ti_{49.7}Cu_{5}$ | 57.15 | 25.2 | 2535-4551 | . |
| . | $Ni_{45.5}Ti_{49.5}Cu_{5}$ | 48.4 | 24.3 | 2445-4388 | . |
| . | $Ni_{45.7}Ti_{49.3}Cu_{5}$ | 30.05 | 21.7 | 2183-3919 | . |
| . | $Ni_{45.9}Ti_{49.1}Cu_{5}$ | 10.45 | 14.3 | 1438-2582 | . |
| . | $Ni_{46}Ti_{49.1}Cu_{5}$ | 2.7 | 16.8 | 1690-3034 | . |
| . | $Ni_{46.1}Ti_{48.9}Cu_{5}$ | -7 | 19.4 | 1952-3503 | . |
| . | $Ni_{46.2}Ti_{48.8}Cu_{5}$ | -21.4 | 14 | 1408-2528 | . |
| . | $Ni_{47.5}Ti_{50}Cu_{2.5}$ | 94.2 | 31 | 3119-5598 | . |
| . | $Ni_{42.5}Ti_{50}Cu_{7.5}$ | 76.8 | 27.9 | 2807-5038 | . |
| . | $Ni_{49.8}Ti_{48.2}Hf_{2}$ | 103.5 | 29.3 | 2948-5291 | . |
| . | $Ni_{49.8}Ti_{46.2}Hf_{4}$ | 114.65 | 29.1 | 2928-5255 | . |
| . | $Ni_{49.8}Ti_{44.2}Hf_{6}$ | 128.35 | 29.1 | 2928-5255 | . |
| . | $Ni_{49.8}Ti_{42.2}Hf_{8}$ | 160.2 | 30.3 | 3048-5472 | . |
| . | $Ni_{49.8}Ti_{40.2}Hf_{10}$ | 178.6 | 29 | 2917-5237 | . |
| . | $Ni_{49.8}Ti_{39.2}Hf_{11}$ | 189.45 | 29.3 | 2948-5291 | . |
| . | $Ni_{49.8}Ti_{35.2}Hf_{15}$ | 248.45 | 31.3 | 3149-5652 | . |
| . | $Ni_{49.8}Ti_{30.2}Hf_{20}$ | 330.2 | 35.1 | 3531-6339 | . |
| . | $Ni_{45}Ti_{50}Pd_{5}$ | 72.2 | 25.8 | 2595-4659 | . |
| . | $Ni_{43}Ti_{50}Pd_{7}$ | 66.2 | 23.8 | 2394-4298 | . |
| . | $Ni_{41}Ti_{50}Pd_{9}$ | 63.25 | 18.4 | 1851-3323 | . |
| . | $Ni_{30}Ti_{50}Pd_{11}$ | 68.15 | 14.5 | 1458-2618 | . |
| . | $Ni_{37}Ti_{50}Pd_{13}$ | 86.05 | 16.4 | 1650-2961 | . |
| . | $Ni_{35}Ti_{50}Pd_{15}$ | 99.35 | 17.7 | 1780-3196 | . |
| . | $Ni_{33}Ti_{50}Pd_{17}$ | 118.45 | 18.7 | 1881-3377 | . |
| . | $Ni_{29}Ti_{50}Pd_{21}$ | 160.45 | 21.6 | 2173-3900 | . |
| . | $Ni_{20}Ti_{50}Pd_{30}$ | 267.85 | 29.7 | 2988-5363 | . |
| . | $Ni_{49.5}Ti_{49.5}Zr_{1}$ | 100.6 | 30.6 | 3078-5526 | . |
| . | $Ni_{49.5}Ti_{47.5}Zr_{3}$ | 102.7 | 23.3 | 2344-4207 | . |
| . | $Ni_{49.5}Ti_{45.5}Z_{r5}$ | 104 | 27 | 2716-4876 | . |
| . | $Ni_{49.5}Ti_{40.5}Zr_{10}$ | 142 | 27.3 | 2746-4930 | . |
| . | $Ni_{49.5}Ti_{35.5}Zr_{15}$ | 221.65 | 28.8 | 2897-5201 | . |
| . | $Ni_{49.5}Ti_{30.5}Zr_{20}$ | 322.55 | 35 | 3521-6321 | . |
| . | $Ni_{28.5}Ti_{50.5}Pt_{21}$ | 330.85 | 22.4 | 2606-4677 | . |
| . | $Ni_{43.5}Ti_{50}Cu_{5}Co_{1.5}$ | 53.85 | 25.9 | 2606-4677 | . |
| . | $Ni_{42}Ti_{50}Cu_{5}Co_{3}$ | 30.2 | 23.8 | 2394-4298 | . |
| . | $Ni_{40}Ti_{50}Cu_{5}Co_{5}$ | -5.35 | 21.6 | 2173-3900 | . |
| . | $Ni_{42.9}Ti_{50}Cu_{5}Pd_{2.1}$ | 66.35 | 14.9 | 1499-2690 | . |
| . | $Ni_{41.9}Ti_{50}Cu_{5}Pd_{3.1}$ | 73.1 | 16 | 1609-2889 | . |
| . | $Ni_{39.2}Ti_{50}Cu_{5}Pd_{5.8}$ | 67.45 | 14.2 | 1428-2564 | . |
| . | $Ni_{38.3}Ti_{50}Cu_{5}Pd_{6.7}$ | 65.65 | 13.6 | 1368-2456 | . |
| . | $Ni_{32.2}Ti_{50}Cu_{5}Pd_{12.8}$ | 59.85 | 12.3 | 1237-2221 | . |
| **SL-PCM** | Paraffin | 64 | 173.6 | 23 | 7 |
| **SS-PCM** | Neopentyl glycol (NPG) | 44 | 131 | 32 | 8 |
| . | $C_{12}Cu$ | 58.15 | 63.8 | 16 | . |
| . | $C_{14}Cu$ | 74.35 | 163.99 | 45 | . |
| . | $C_{15}Cu$ | 80.05 | 87.8 | 25 | . |
| . | Pentaglycerine (PG) | 85 | 193 | 52 | . |
| . | Form stable HDPE | 133 | 188 | 42 | . |
| . | Pentaerythritol (PE) | 185 | 303 | 97 | . |
| . | PlusICE X25 | 25 | 110 | 42 | 9 |
| . | PlusICE X30 | 30 | 105 | 40 | . |
| . | PlusICE X40 | 40 | 125 | 47 | . |
| . | PlusICE X55 | 55 | 115 | 44 | . |
| . | PlusICE X70 | 70 | 125 | 49 | . |
| . | PlusICE X80 | 80 | 140 | 60 | . |
| . | PlusICE X90 | 90 | 135 | 58 | . |
| . | PlusICE X95 | 95 | 140 | 61 | . |
| . | PlusICE X120 | 120 | 180 | 81 | . |
| . | PlusICE X130 | 130 | 260 | 120 | . |
| . | PlusICE X165 | 165 | 230 | 108 | . |
| . | PlusICE X180 | 180 | 280 | 134 | . |

## REFERENCES

(1) Cahill, D.G.; Analysis of heat flow in layered structures for time-domain

thermoreflectance. *Review of Scientific Instruments*. **2004**, 75, 5119-5122.

(2) Schmidt, A.J.; Cheaito, R.; Chiesa, M. A frequency-domain thermoreflectance method for the characterization of thermal properties. *Review of Scientific Instruments*. **2009**, 80, 094901.

(3) Boyer, R.; Collings, E.W.; Welsch, G. *Materials Properties Handbook: Titanium Alloys*. ASM International, Materials Park, USA, **1994**.

(4) Otubo, J.; Rigo, O.; Coelho, A.; Neto, C.; Mei, P. The influence of carbon and oxygen content on the martensitic transformation temperatures and enthalpies of NiTi shape memory alloy. *Mater. Sci. Eng.* **2008**, *481, 639-642*.

(5) Frenzel, J.; Wieczorek, A.; Opahle, I.; MaaB, B.; Drautz, R.; Eggeler, G. On the effect of alloy composition on martensite start temperatures and latent heats in Ni-Ti based shape memory alloys. *Acta Mater.* **2015**, *90, 213-231*.

(6) Chirtoc, M.; Gibkes, J.; Wernhardt, R.; Pelzl, J.; Wieck, A. Temperature-dependent quantitative 3$\omega$ scanning thermal microscopy: Local thermal conductivity changes in NiTi microstructures induced by martensite-austenite phase transition. *Review of Scientific Instruments*. **2008**, 79, 093703.

(7) Shao, L.; Raghavan, A.; Kim, G.; Emurian, L. Figure of merit for phase change materials used in thermal management. *Int. J. of Heat and Mass Trans.* **2016**, *101, 764-771*.

(8) Jankowski, N.; McCluskey, F. A review of phase change materials for vehicle component thermal buffering. *App. Energy.* **2014**, *113, 1525-1561*.

(9) P.P. Limited. PlusICE Phase Change Materials. [online] available:

http://www.pcmproducts.net/files/PlusICE%20Range-2013.pdf.